\documentstyle[psfigs]{mn}
\newcommand{\ergcms}{erg\,cm$^{-2}$\,s$^{-1}$}

\begin{document}

\title{The role of nonthermal electrons in the optical continuum of
  stellar flares}

\author[M.~D.~Ding and C.~Fang]
       {M.~D.~Ding and C.~Fang \\
        Department of Astronomy, Nanjing University, Nanjing 210093, China}

\date{Accepted 2000 . Received 2000 ; in original form 2000 }

\pagerange{\pageref{firstpage}--\pageref{lastpage}}
\pubyear{2000}

\maketitle

\label{firstpage}

\begin{abstract}

The continuum emission of stellar flares in UV and visible bands can be
enhanced by two or even three orders of magnitude relative to the quiescent
level and is usually characterised by a blue colour. Thermal atmospheric
models are difficult to reproduce all these spectral features. If the
flaring process involves the acceleration of energetic electrons which then
precipitate downwards to heat the lower atmosphere, collisional excitation
and ionisation of ambient hydrogen atoms by these nonthermal electrons
could be important in powering the continuum emission. To explore such a
possibility, we compute the continuum spectra from an atmospheric model for
a dMe star, AD Leo, at its quiescent state, when considering the nonthermal
effects by precipitating electron beams. The results show that if the
electron beam has an energy flux large enough (for example, ${\cal F}_{1}
\sim 10^{12}$ \ergcms), the $U$ band brightening and, in
particular, the $U-B$ colour are roughly comparable with observed values
for a typical large flare. Moreover, for electron beams with a moderate
energy flux ${\cal F}_{1}\la 10^{11}$ \ergcms, a decrease
of the emission at the Paschen continuum appears. This can explain at
least partly the continuum dimming observed in some stellar flares.
Adopting an atmospheric model for the flaring state can further raise the
continuum flux but it yields a spectral colour incomparable with
observations. This implies that the nonthermal effects may play the chief
role in powering the continuum emission in some stellar flares.

\end{abstract}


\section{Introduction}

Flares have been detected in stars of various types. In particular, very
energetic flares are likely to occur in dKe/dMe stars. The most energetic
stellar flare can release an energy as large as $\sim 10^{35}$ erg in
total, three orders of magnitude greater than that in the biggest flare
on the Sun (see e.g. Pagano et al. 1997). Multiwavelength observing
campaigns have been made to detect flare emissions in different wavelength
regions like the soft X-ray, EUV, optical, and radio and then to quantify
their relationships (e.g. Hawley \& Pettersen 1991; Hawley et al. 1995;
van den Oord et al. 1996). Though, there are still few hard X-ray
observations of stellar flares.

In spite of many different aspects between stellar and solar flares
(Haisch, Strong \& Rodon\`o 1991), they may share similarities in some
basic physical
processes, such as the energy release process, heating of the atmosphere,
and origin of the continuum and line emissions. The well-known solar
flare model assumes that a flare is produced through the reconnection of
magnetic field lines, leading to the formation of a current sheet where
tearing mode instability occurs. This process releases a large amount of
thermal energy and results in an acceleration of electrons and/or
protons. Accelerated electrons then impact the lower atmosphere, producing
hard X-ray emission through bremsstrahlung, and causing an enhancement of
optical continuum and line emissions (e.g. Neidig et al. 1993).
Meanwhile, the corona accumulates material evaporated from the
chromosphere when heated by nonthermal electrons and the soft X-ray flux
gradually increases. The `Neupert effect', assuming that the soft X-ray
flux is proportional to the time integral of hard X-ray flux, is found
valid not only for solar flares (Dennis \& Zarro 1993), but also for
stellar flares if using optical data as a proxy for hard X-ray emission
(Hawley et al. 1995).

The UV and optical continuum in stellar flares is found to exhibit
several emission features. First, the continuum is usually characterised
by a blue colour. Pagano et al. (1997) found in an intense flare a very
blue emission ($U-B\approx -2.5$), implying a significant enhancement of
the Balmer continuum and a flare site close to the stellar limb. A blue
continuum was also observed by Hawley \& Fisher (1992), which was
reproduced from atmospheric models proposed by Mauas \& Falchi (1996),
taking into account the merging of higher Balmer lines. Second, some
flares are preceded by a dimming in their continuum emission (e.g.
Hawley et al. 1995).
Several models have been proposed to explain such negative flares.
Gurzadyan (1980) proposed inverse Compton up-conversion of `red'
photons to the $U$ band by extremely energetic electrons. However,
this model demands a total energy ($\sim 10^{40}$ erg) of energetic
electrons which is unrealistically large (H\'enoux et al. 1990).
Mullan (1975) suggested that the dips in the red light curve can be
produced by H$\alpha$ absorption. At present, the most favourable model
is the H$^{-}$ absorption model of Grinin (1983), which suggests that
an energy input to the lower atmosphere leads to an enhanced H$^{-}$
opacity and then reduces the photospheric continuum radiation.

The situation is more or less similar in solar white-light flares (WLFs),
those flares emitting an enhanced optical continuum. The majority of
solar WLFs exhibit a pronounced Balmer jump while a small fraction shows
no jump, which are classified as type I and type II WLFs, respectively
(Machado et al. 1986; Fang \& Ding 1995). Recombination of hydrogen atoms
in the chromosphere is responsible for the continuum emission in the
first case while the H$^{-}$ emission plays the main role in the second
case. Besides, a negative continuum contrast is also theoretically
predicted at the very beginning of flare occurrence (H\'enoux et al. 1990;
Ding \& Fang 1996). Contrary to the stellar case, a negative solar WLF,
the so-called black-light flare (BLF), is hard to observe due to its
small amplitude of dimming and its short duration. In fact, no direct
observational evidence has proved its existence (van Driel-Gesztelyi
et al. 1994).

Since the lower atmosphere of flares is very likely to be heated through
bombardment of energetic electrons, the nonthermal collisional excitation
and ionisation of hydrogen atoms can be important in increasing the
continuum opacity and source function (H\'enoux, Fang \& Gan 1993).
The emergent
continuum emission is then greatly enhanced and characterised by a Balmer
jump. The continuum contrast depends on the flux and spectrum of the
electron beam and also on the flare site (Ding \& Fang 1996).

The atmospheric structure and, in particular, the parameters of the
electron beam in stellar flares can be quite different from that in a
solar flare. Thus, it is still unclear and needs to be quantified to what
extent the nonthermal effects can alter the continuum emission in stellar
flares. The purpose of this paper is to explore this question. To do so,
we employ atmospheric models typical for dMe stars and do non-LTE
calculations in the presence of an electron beam. The results show the
role of nonthermal effects on the optical continuum and are useful in the
spectroscopy of stellar flares.

\section{Computational Method}

\subsection{Atmospheric models}

In last decades, attempts have been made to model the atmosphere of
stars of different spectral types. Here, we restrict our attention to
the work on late-type stars, in particular, M dwarfs, which have a
lower effective temperature. The reason is that in such stars, the
afore-mentioned nonthermal effects during flares may be relatively
more significant.

Some early atmospheric models of dM or dMe stars were built on limited
spectral data. Thus, their chromospheric structures are usually very
schematic (e.g. Cram \& Mullan 1979). Multi-line spectral data are
needed for a better modelling of the stellar atmosphere. Recently,
after constructing grids of atmospheric models and examining their
different spectral features, methods for probing the atmosphere are
being developed which use spectral lines including H\,{\sc i} Ly$\alpha$,
H$\alpha$ (H$\beta$), Ca\,{\sc ii} H (K), Mg\,{\sc ii} h (k), and
Na\,{\sc i} D lines etc. as well as continua (e.g. Doyle et al. 1994;
Houdebine \& Doyle 1994; Houdebine, Doyle \& Ko\'scielecki 1995;
Houdebine et al. 1996; Houdebine \& Stempels 1997; Short \& Doyle 1998).

Mauas \& Falchi (1994) presented a semi-empirical model for the dMe star
AD Leo, which can reproduce both the continuum and most spectral lines
from observations. Their model is elaborate especially in the
chromospheric structure, though, there may be some uncertainties in the
temperature minimum region, as they stated. Mauas \& Falchi (1996)
further constructed semi-empirical models for a large flare on AD Leo
observed by Hawley \& Pettersen (1991). They computed three models
corresponding to different filling factors. Compared to the quiescent
atmosphere, these models indicate a strong heating in both the
chromosphere and the photosphere, which is quite different from the case
of a solar flare. In particular, Mauas \& Falchi (1996) discussed the
possible role of high energy electrons on heating the deeper atmosphere.

\subsection{Nonthermal excitation and ionisation rates of hydrogen by
electron beams}

Here we assume that the basic scenario for solar flares is still ture
for stellar flares, that is, flares originate in stellar coronae and
involve the production of a beam of high energy electrons, streaming
downward to heat the lower atmosphere. In the cold target approximation,
the rate of energy loss for beam electrons with column density $N$ has
the form (Emslie 1978; H\'enoux et al. 1993):
\begin{equation}
\frac{{\mathrm d}E}{{\mathrm d}N} = -\frac{K}{\mu E}\gamma
  = -\frac{K}{\mu E} [\xi\Lambda+(1-\xi)\Lambda'],
\label{eq1}
\end{equation}
where $K=2\pi e^{4}$. In equation (\ref{eq1}), $\mu$ is the
cosine of the pitch angle, $\xi$ is the ionisation degree of the target,
$\Lambda$ is the Coulomb logarithm due to collisions with target electrons
and protons, while $\Lambda'$ the one for inelastic collisions with
target atoms. The latter corresponds to the energy deposit contributed
to the nonthermal excitation and ionisation of target atoms. The rate of
energy deposit at column density $N$ through inelastic collisions is
written as (Emslie 1978; Chambe \& H\'enoux 1979):
\begin{equation}
\Phi = \frac{{\mathrm d}E}{{\mathrm d}t} =
  (1-\xi)n_{\mathrm H}\Lambda' K
  \int^{\infty}_{E_{N}} \frac{F_{0}(E) {\mathrm d}E}
  {E(1-E^{2}_{N}/E^{2})^{\frac{2+\beta}{4+\beta}}},
\label{eq2}
\end{equation}
where
\begin{equation}
E_{N}=\left[ \left(2+\frac{\beta}{2}\right)
  \frac{\gamma KN}{\mu_{0}}\right]^{\frac{1}{2}}
\end{equation}
is the minimum energy of electrons that can penetrate to $N$. The cosine
of the initial pitch angle, $\mu_{0}$, is taken to be unity throughout
the paper. The expression for $\beta$ can be found in Emslie (1978).
Knowing little about the energy distribution of the beam electrons,
we assume a power law for the initial flux spectrum with a low-energy
cut-off:
\begin{equation}
F_{0}(E)=\left\{ \begin{array}{ll}
                  AE^{-\delta}, & E>E_{1},\\
                  0,            & E<E_{1},
                 \end{array}
         \right.
\end{equation}
where $A=(\delta-2){\cal F}_{1}E^{\delta -2}_{1}$. ${\cal F}_{1}$ is the
energy flux of electrons. Equation (\ref{eq2}) then reduces to
\begin{equation}
\Phi = \frac{1}{2} (1-\xi)n_{\mathrm H}\Lambda' K(\delta -2)
  \frac{{\cal F}_{1}}{E^{2}_{1}} \left( \frac{N}{N_{1}} \right)
  ^{-\frac{\delta}{2}} \int^{u_{1}}_{0}
  \frac{u^{\frac{\delta}{2}-1}{\mathrm d}u}
  {(1-u)^{\frac{2+\beta}{4+\beta}}},
\end{equation}
where
\begin{equation}
N_{1}=\frac{\mu_{0}E^{2}_{1}}{\left(2+\beta/2\right) \gamma K}
\end{equation}
is the column density penetrated by particles with an initial energy
$E_{1}$, and
\begin{equation}
u_{1}=\left\{ \begin{array}{ll}
                1,        & N>N_{1},\\
                N/N_{1},  & N<N_{1}.
              \end{array}
      \right.
\end{equation}

In the present calculations, we adopt an atomic model for hydrogen with
four bound levels plus continuum. According to Fang, H\'enoux \& Gan
(1993), the nonthermal collisional excitation rates due to bombarding
electrons from the ground state are derived as
\begin{equation}
  C^{\mathrm B}_{12}\simeq 2.94\times 10^{10}\Phi/n_{1},
\end{equation}
\begin{equation}
  C^{\mathrm B}_{13}\simeq 5.35\times 10^{9}\Phi/n_{1},
\end{equation}
and
\begin{equation}
  C^{\mathrm B}_{14}\simeq 1.91\times 10^{9}\Phi/n_{1},
\end{equation}
and the nonthermal ionisation rate is
\begin{equation}
C^{\mathrm B}_{\mathrm 1c}\simeq 1.73\times 10^{10}\Phi/n_{1}.
\end{equation}
These rates are then incorporated into the statistical equilibrium
equations, which are solved iteratively together with the radiative
transfer equation for a specific atmospheric model. Finally the continuum
flux at the stellar surface is computed as
\begin{equation}
F_{\lambda}=2\pi\int^{1}_{0} I_{\lambda}(\mu) \mu {\mathrm d}\mu =
  2\pi\int^{1}_{0}{\mathrm d}\mu \int^{\infty}_{0}
  S_{\lambda}{\mathrm e}^{-\tau_{\lambda}/\mu}{\mathrm d}\tau_{\lambda}.
\end{equation}

\section{Results and Discussions}

Using the above method, we have studied the role of nonthermal electrons
on the continuum emission in stellar flares. Since the flaring atmosphere
undergoes a rapid evolution during the heating process, we consider two
quasi-static cases (marked as A and B), an initial cool atmosphere and a
heated atmosphere, which are assumed to be represented respectively by the
atmospheric model for the quiescent state (Mauas \& Falchi 1994) and that
for a flaring state (Mauas \& Falchi 1996, their model B) of the M dwarf
AD Leo. This is reasonable as the radiative time scale is much shorter
than the hydrodynamic time scale and our main interest is the radiative
output. These two cases can be regarded as the very early and the maximum
phases of flares respectively. The continuum emission is then computed for
the two models when considering the nonthermal collisional effects by
electron beams precipitating from the corona.

\begin{figure}
\centerline{\psfig{file=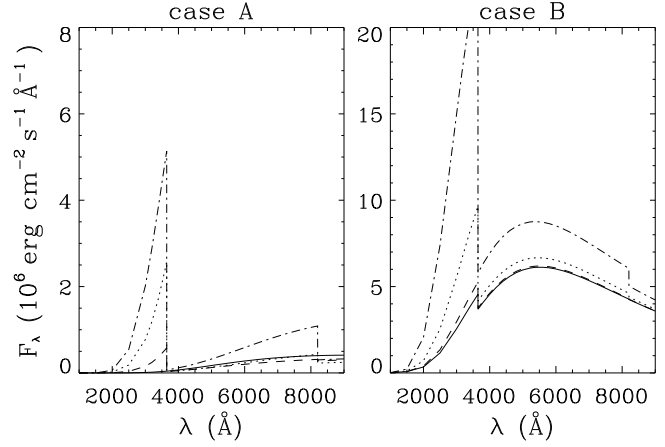, angle=0, width=8.8cm}}
\caption[]{Continuum spectra computed for a quiescent model (case A,
left panel) and a flare model (case B, right panel) when considering the
nonthermal collisional effects by precipitating electron beams. The
energy fluxes, ${\cal F}_{1}$, of the beams are 0 (solid line), 10$^{10}$
(dashed line), 10$^{11}$ (dotted line), and 10$^{12}$ (dash-dotted line)
\ergcms. For all cases, the low-energy cut-off, $E_{1}$,
is 20 keV, and the power index, $\delta$, is 3.}
\label{fig1}
\end{figure}

As an example, Fig.~\ref{fig1} plots the continuum spectra computed for
both models in the presence of electron beams with a low-energy cut-off
$E_{1}=20$ keV and a power index $\delta=3$ but different energy fluxes
${\cal F}_{1}=0$, 10$^{10}$, 10$^{11}$, and 10$^{12}$ \ergcms.
It clearly shows that the bombardment of an electron beam
leads to a dramatic increase of the continuum emission in the Balmer
continuum region ($\lambda < 3646$ \AA). The flux at the Paschen
continuum region ($\lambda < 8204$ \AA) changes to a relatively smaller
extent. Therefore, a strong Balmer jump appears in the computed
spectra which has not been observed (see also, Hawley \& Fisher 1992).
Mauas \& Falchi (1996) proposed that the merging of higher Balmer lines
results in a pseudo-continuum which can smear out the Balmer jump. In
the present case, the nonthermal electrons are expected to strengthen the
Balmer lines as well (Fang et al. 1993). Thus, the merging effect could be
even more important here. In addition, the Doppler effect due to velocity
fields of several hundreds of km s$^{-1}$ possibly existing in the
flaring atmosphere (e.g. Houdebine et al. 1993), and the line
blanketing effect, which is not considered here, can also help to
lessen the amplitude of the Balmer jump.

\subsection{Enhanced continuum emission and the spectral colour}

To make a comparison with observations, we extract the fluxes at
$\lambda=3600$ and 4400 \AA, which lie in the centre of the
Johnson $U$ and $B$ bands, respectively, and then compute the
following parameters,
\begin{equation}
\Delta U'=-2.5\log [F_{\mathrm 3600,f}/F_{\mathrm 3600,q}],
\end{equation}
\begin{equation}
\Delta B'=-2.5\log [F_{\mathrm 4400,f}/F_{\mathrm 4400,q}],
\end{equation}
and
\begin{equation}
\Delta (U'-B')=\Delta U'-\Delta B'.
\end{equation}
In the above equations, a subscript `f' refers to the flaring status
(a quiescent/flare atmospheric model with a precipitating electron beam),
while `q' to the pre-flare status (a quiescent atmospheric model only).
A prime means that the parameters are defined at fixed wavelength points
instead of broad bands usually used in observations. Thus, the results
can only be considered as exploratory.

\begin{figure}
\centerline{\psfig{file=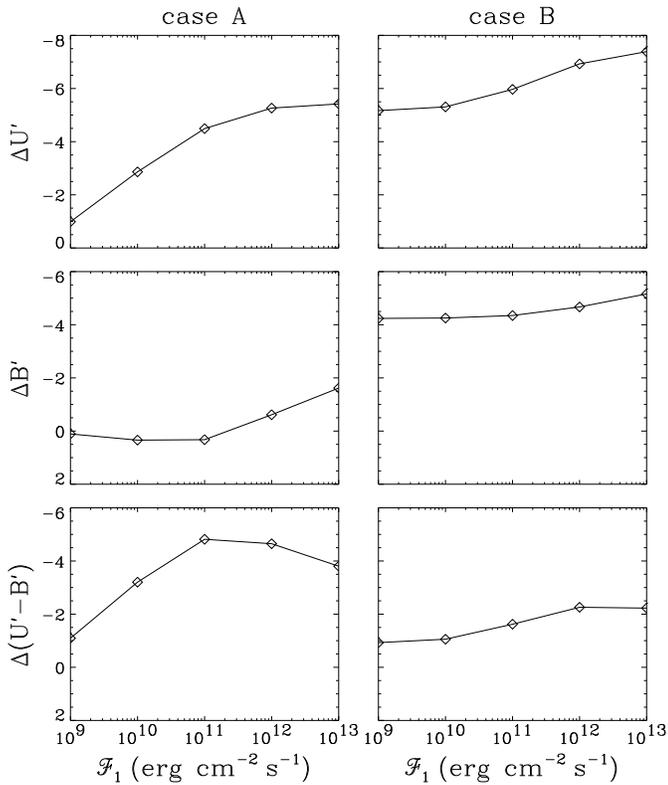, angle=0, width=8.8cm}}
\caption[]{Computed parameters of $\Delta U'$, $\Delta B'$, and
$\Delta(U'-B')$ in dependence of the energy flux of the precipitating
electron beam. For all cases, the low-energy cut-off, $E_{1}$, is 20 keV,
and the power index, $\delta$, is 3.}
\label{fig2}
\end{figure}

Fig.~\ref{fig2} displays the computed parameters versus the electron
beam fluxes. One finds that in the case of a large beam flux, pure
nonthermal effects can help to raise the emission at $\lambda=3600$ \AA\
by two orders of magnitude. For example, if ${\cal F}_{1}$ reaches
$\sim 10^{12}$ \ergcms, one obtains an enhancement corresponding to
$\Delta U'\sim -5.3$ (case A). Adopting a heated atmospheric model
with the same electron beam yields further $\Delta U'\sim -6.9$
(case B). This means that in the $U$ band, the nonthermal effects could
dominate over the effect of temperature rise (the thermal effect) in
producing the continuum emission. Pagano et al. (1997) reported on a
large flare occurring on a dMe star G 102-21, which had a $U$ band
enhancement of $\approx 7.3$ mag. Taking into account the different
pre-flare atmospheric conditions, we conclude that the $U$ band emission
in this flare can be qualitatively explained in terms of nonthermal
effects.

Under the current framework that the electron beam originates in the
corona, it is conceivable that the Paschen continuum is less affected
by the electron beam since it is formed in deeper layers than the Balmer
continuum. Judging from Fig.~\ref{fig2}, the thermal contribution to
the emission in the $B$ band may remain to prevail over that due to
nonthermal effects. For an electron beam with ${\cal F}_{1}\sim 10^{12}$
\ergcms, $\Delta B'$ is computed to be $\sim -0.6$ and $-4.7$
in cases A and B, respectively. The latter value is closer to the
observations by Pagano et al. (1997), who obtained a $B$ band
enhancement of $\approx 3.9$ mag.

Due to the greatly enhanced emission in the $U$ band, the computed
spectral colour appears very blue. The colour change, $\Delta (U'-B')$,
is computed to show the difference of relative enhancements at $U'$ and
$B'$ wavelengths. In case A, this parameter varies in a broad range.
The bluest colour, $\Delta (U'-B')\sim -4.8$, appears when
${\cal F}_{1}\sim 10^{11}$ \ergcms. The colour turns
less blue if further increasing the values of ${\cal F}_{1}$. Thus, the
observed colour change, $\Delta(U-B)\approx -3.4$ (Pagano et al. 1997),
is easy to be reproduced. In case B, however, the spectra are
contaminated by the thermal effect, the bluest colour of which is shown
to be $\Delta (U'-B')\sim -2.3$ for electron beams considered here.
The different results in these two cases imply that the observed blue
colour is very likely of nonthermal origin. A pure thermal model is
practically hard to reproduce such a blue colour.

Finally, it is worth noting that an electron beam can produce a return
current instability when its flux exceeds some critical value. According
to the formula derived by Aboudarham \& H\'enoux (1986), we have found
that electron beams with ${\cal F}_{1}\sim 10^{12}$ \ergcms are stable
in the corona of the flare model while they are only marginally stable in
the corona of the quiescent model. Hence, a favourable situation for the
existence of such a large energy flux includes a relatively higher
preflare coronal pressure and a harder spectrum of the electron beam
(a smaller $\delta$ and/or a larger $E_{1}$).

\subsection{The cause of a continuum dimming}

A continuum dimming is produced at the Paschen continuum ($B'$ and $V'$
wavelengths) in the case of a cool atmosphere (case A) bombarded by an
electron beam with a moderate energy flux (${\cal F}_{1}\la 10^{11}$
\ergcms). However, such a darkening switchs quickly to a
brightening with increasing values of ${\cal F}_{1}$. In the present
models, we cannot produce a dimming at the Balmer continuum.

\begin{figure}
\centerline{\psfig{file=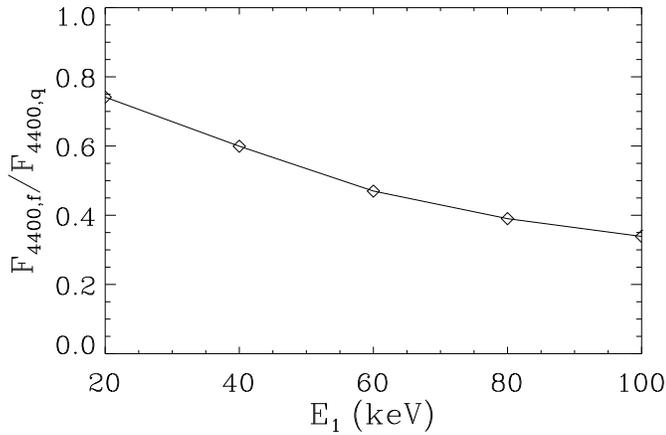, angle=0, width=8.8cm}}
\caption[]{Computed continuum dimming,
$F_{\mathrm 4400,f}/F_{\mathrm 4400,q}$, in dependence of the low-energy
cut-off of the precipitating electron beam. For all cases, the energy
flux, ${\cal F}_{1}$, is 10$^{11}$ \ergcms, and the power
index, $\delta$, is 3.}
\label{fig3}
\end{figure}

The continuum dimming results from attenuation of the photospheric
radiation by an enhanced opacity in the lower chromosphere and the
temperature minimum region. Nonthermal ionisation of hydrogen atoms
overpopulates the ambient electrons, leading to an enhanced H$^{-}$
opacity. Therefore, the magnitude of the depression depends sensitively
on the electron energy, which determines the electron penetration depth.
Fig.~\ref{fig3} illustrates the effect on the continuum dimming at
$\lambda=4400$ \AA\ of varying the low-energy cut-off, while keeping
the energy flux of the electron beams unchanged. It clearly shows a
positive dependence of the magnitude of the dimming on the electron
energy.

In observations, the $U$ band comprises a part of the Paschen continuum.
Thus, we think that the observed continuum dimming could be related to
an electron beam bombarding a relatively cooler atmosphere. This is
likely to occur in the early phase of flares. In conclusion, the
nonthermal effects could at least partly, if not fully, account for the
continuum dimming observed in some stellar flares (e.g. Hawley et al.
1995).

\begin{figure}
\centerline{\psfig{file=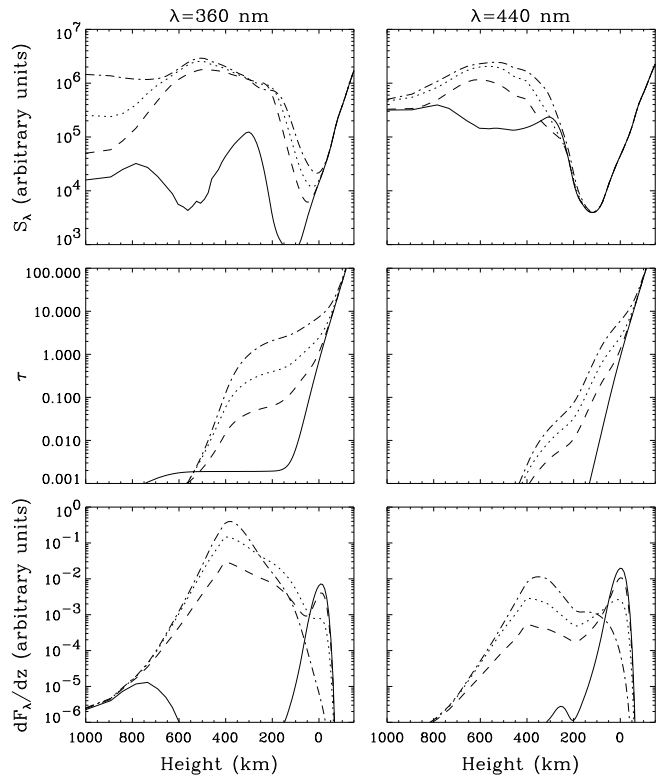, angle=0, width=8.8cm}}
\caption[]{Height distributions of the source function, optical depth and
the contribution function at $\lambda=3600$, and 4400 \AA\ for case A. The
energy fluxes, ${\cal F}_{1}$, of the electron beams are 0 (solid line),
10$^{10}$ (dashed line), 10$^{11}$ (dotted line), and 10$^{12}$
(dash-dotted line) \ergcms. For all cases, the low-energy
cut-off, $E_{1}$, is 20 keV, and the power index, $\delta$, is 3.}
\label{fig4}
\end{figure}

\subsection{Interpretations of the results}

To understand more clearly the above results, we give in Fig.~\ref{fig4}
the height distributions of the source function, optical depth and the
contribution function per unit geometrical depth,
\begin{equation}
{\mathrm d}F_{\lambda}/{\mathrm d}z=2\pi\int^{1}_{0} j_{\lambda}
  {\mathrm e}^{-\tau_{\lambda}/\mu} {\mathrm d}\mu,
\end{equation}
at $\lambda=3600$, and 4400 \AA\ for case A. It shows that the nonthermal
effects can raise both the source function and the opacity to a large
extent in the chromosphere. Thus, the chromospheric contribution to the
emergent flux gradually increases while the photospheric contribution
decreases with increasing electron beam flux. Whether the continuum
brightens or darkens is then determined by the competition of these
two factors.

\section{Conclusions}

If stellar flares result from the reconnection of magnetic fields,
analogously to solar flares, electrons and/or protons could be accelerated
at the reconnection site and then precipitate into the lower atmosphere.
The nonthermal collisional excitation and ionisation of ambient hydrogen
atoms by these electrons lead to an enhancement of the source function
and opacity in the atmosphere. The chromospheric contribution to the
continuum emission rises rapidly, producing a spectrum with a very blue
colour. By employing an atmospheric model for an M dwarf star, AD Leo,
at quiescent state (Mauas \& Falchi 1994), we have computed the continuum
spectra for various electron beams bombarding the atmosphere from the
corona. The results show that an intense electron beam with an energy
flux, for example, ${\cal F}_{1}\sim 10^{12}$ \ergcms\ can
produce a continuum flux at $\lambda=3600$ \AA\ (Balmer continuum) two
orders of magnitude greater than the quiescent value. The $U$ band
brightening and, in particular, the $U-B$ colour are roughly comparable
with observed values for a typical large flare. Adopting an atmospheric
model for the flaring state can further raise the continuum flux and
can better account for the $B$ band enhancement. However, it yields a
spectral colour incomparable with observations. This implies that the
nonthermal effects may play the chief role in powering the continuum
emission in some stellar flares.

Computations also predict a continuum dimming at the Paschen continuum,
due to attenuation of the photospheric radiation by an enhanced opacity in
the lower chromosphere and the temperature minimum region. This happens
in a cool atmosphere bombarded by an electron beam with a moderate energy
flux (${\cal F}_{1}\la 10^{11}$ \ergcms). This is likely
to fit the conditions in the very early phase of flares. The magnitude of
the dimming depends positively on the electron energy. We argue that the
continuum dips found in observations (e.g. Hawley et al. 1995) could
possibly be related to nonthermal electrons of higher energies.

\section*{acknowledgements}
The authors are very grateful to the referee, Prof. J. C. Brown, for
his valuable comments on the manuscript. This work was supported by the
Research Fund for the Doctoral Programme of Higher Education and by a
grant from the National Natural Science Foundation of China.

\label{lastpage}

\end{document}